\documentclass[aps,twocolumn,showpacs,preprintnumbers,amsmath,amssymb,floatfix]{revtex4-2}
\usepackage{graphicx}
\usepackage{epsfig}
\begin{document}
\def\be{\begin{equation}}
\def\ee{\end{equation}}
\def\bearr{\begin{eqnarray}}
\def\eearr{\end{eqnarray}}
\def\tc{$T_c~$}
\def\half{$\rm \frac{1}{2}$~}
\def\Moire{$\rm Moir\'e $~}
\def\tperp{t$_\perp~$}
\def\jperp{J$_\perp~$}
\def\utilde{$\tilde{\rm U}~$}

\title{Metastable Kitaev Spin Liquids in Isotropic Quantum Heisenberg Magnets}

\author{G. Baskaran}

\affiliation
{The Institute of Mathematical Sciences, C.I.T. Campus, Chennai 600 113,\\
Department of Physics, Indian Institute of Technology, Chennai 600 036, India \&\\
Perimeter Institute for Theoretical Physics, Waterloo, ON N2L 2Y5, Canada}

\begin{abstract} 
Metastable states with surprising properties abound in Hilbert space. We study unfrustrated isotropic spin-\half Heisenberg models in honeycomb lattice and find emergence of \textit{metastable Kitaev spin liquids having a 2-spin nematic long range order}, via spontaneous symmetry breaking. Decomposition of isotropic Heisenberg Hamiltonian $H^H$ into an exact sum of 3 noncommuting (permuted) Kitaev Hamiltonians, $H^H$ = $H^{K}_{\rm xyz}+H^{K}_{\rm yzx}+H^{K}_{\rm zxy},$ helps us build a degenerate \textit{manifold of flux free metastable Kitaev spin liquid vacua} and vector Fermionic (Goldstone like) collective modes. We introduce a method, \textit{symmetric decomposition of Hamiltonians}, which might help craft \textit{designer metalstable phases}. It is likely that  small Kitaev interactions present in Jackeli-Khaliullin-Kitaev materials, with dominant Heisenberg couplings, bring in metastable Kitaev spin liquid features in real experiments. Present work opens possibilities of performing quantum computation and other tasks, using exotic quasiparticles and exotic metastable states, present in nonexotic real systems.
\end{abstract} 
\maketitle

\section*{I. Introduction}
In the incomprehensibly vast many body Hilbert space of model Hamiltonians, novel metastable states, manybody localization, quantum scars, Hilbert space fragmentation, time crystals, quantum orders and phenomena, beyond the familiar ground realities exist \cite{MBL,QScar,HSFrag,TXtl}. Model Hamiltonians, designed to understand certain low energy physics and experiments, are space ships that could take us to new worlds. Such explorations are currently in vogue, thanks to exciting developments in non-equilibrium quantum manybody physics and quantum information science.

Equilibrium stable phases with long range order are well known: scores of crystalline phases of ice, ferro/antiferromagnets, ferro/antiferroelectrics, superfluids, superconductors etc. Metastable phases, which break translational invariance, are built using topological defects in underlying Landau type of order parameters: intertwined network of magnetic domain walls, Skyrmion glasses in magnets, vortex glasses in type-II superconductors, structural glasses, defect networks in crystals etc. Degree of metastability, time scale of relaxation are decided by, among other things, thermal, quantum fluctuations, interaction between defects etc. There are also defect free metastable phases, like supercooled water and possibly supersolid He$^4$.

There is growing interest in going beyond Landau's order parameter paradigm in quantum matter. For example, quantum spin liquids, present in certain frustrated Heisenberg spin models \cite{PWARVB73}, anisotropic Kugel-Khomskii-Kitaev compass models \cite{KK,KitaevHC} etc., harbor exotic neutral Fermion, anyon excitations, emergent gauge fields, quantum orders etc \cite{PWARVB87,FrankW,GBPWAGauge,WenQOrder}. Spin liquids are sources of rich physics and connect us to worlds beyond condensed matter and rich mathematics \cite{SYK,WenBook}. 

Metastable quantum phases with quantum orders or symmetry protected topological orders remain vastly unexplored. We suggest \textit{Symmetric Decomposition of Hamiltonians} (SDH), a route to explore or sculpt designer metastable vacua, having quantum or topological orders. It is illustrated using unfrustrated isotropic Heisenberg spin models. Isotropic spin-\half Heisenberg Hamiltonians with nearest neighbor FM/AFM coupling in honeycomb lattice are known to have spontaneous long range FM/AFM ordered ground states. We build, in these unfrustrated isotropic Heisenberg systems, \textit{metastable Kitaev spin liquid vacua}, exhibiting spontaneous two spin nematic order, quantum order, Goldstone like Fermionic collective modes etc.

Decay of the false Kitaev spin liquid vacuum of an isotropic Heisernberg magnet is a difficult question to analyse; it is beyond the scope of the present article. However, we hypothesize a three stage decay, with a superimposed transient and \textit{time crystal} like local oscillations.

In \textit{Jackeli-Khaliullin-Kitaev materials} \cite{JackeliKhaliullin}, with a dominant Heisernberg coupling, metastable Kitaev spin liquid phenomena are likely to intervene in real experiments, because of an enhanced Kitaev spin liquid susceptibiliy exhibited by isotropic Heisenberg systems. This could explain some of the anomalies and experimental puzzles. 

Our work opens possibilities of performing quantum computation and other tasks, using exotic quasiparticles and exotic metastable states hiding in real nonexotic systems.

\section*{Symmetries of Quantum Spin Liquids}
Anderson proposed spin singlet short range RVB (quantum spin liquid) ground state for  spin-\half isotropic Heisenberg antiferromaget \cite{PWARVB73}, on a frustrated triangular lattice. This paramagnetic quantum spin liquid state has no long range 120$^0$ type single spin order. Further, phase relations among bond singlet configurations can be varied suitably to define family of quantum spin liquids, beyond the simple short range RVB \cite{KRSSutherland}, containing rich variety of quantum orders, topological orders etc. 

It also is possible to have stable paramagnetic quantum spin liquid ground states, which have spontaneous long range two spin nematic orders \cite{Andreev,ColemanPremi,Shannon,Yasir,SWhite}, showing  divergence of corresponding 2 spin susceptibilities, obtained via 4 spin correlation functions. Or three spin scalar chiral orders \cite{Chirality}, exhibiting divergence of corresponding 3 spin susceptibilities, obtained via suitable 6 spin correlation functions. These generalised RVB states have resonating singlets and triplets appearing together in a variety of ways and define a plethora of quantum spin liquids.

Kitaev spin liquid ground state (zero flux sector) is anisotropic in spin space. It has lesser symmetry, from the point of view of global spin symmetry present in isotropic Heisenberg Hamiltonains. However, Kitaev spin liquid has the full symmetry of the Kitaev spin Hamiltonian that has manifest bond anisotropy.  In the present article, we show emergence of such metastable phases and corresponding Goldstone modes, even in the absence of frustration in spin-\half isotropic Heisenberg magnets. (We also find \cite{GBFrancesco} that we could stablize such a spontaneuos symmetry broken Kitaev spin liquid like ground state having nematic order, by incorporating non neighbor Heisernberg couplings in spin-\half, isotropic Heisenberg Hamiltonian in honeycomb lattice).

\section*{FM or AFM Order in Isotropic Heisenberg Hamiltonians}
To study AFM ordered ground state arising via spontaneous symmetry breaking, we traditionally decompose isotropic spin Hamiltonians into solvable pieces. Then start with ground state solution of one of the solvable pieces and include effects of the rest of the pieces using certain approximations. For example,  we decompose non-frustrated nearest neighbor S=\half Heisenberg Hamiltonian $H^H$ in a bipartite lattice into three noncommuting Ising pieces,
\be
H^H = J\sum_{\langle ij \rangle} {\vec{\sigma}}_{i} \cdot {\vec{\sigma}}_{j} =
J\sum_{\langle ij \rangle}({\sigma}^x_{i}{\sigma}^x_{j} + {\sigma}^y_{i}{\sigma}^y_{j} +{\sigma}^z_{i}{\sigma}^z_{j})
\ee
and perform spin wave (stability) analysis, about classical (direct product) ground state of one of the three pieces, say z-axis Ising piece. Holstein-Primakoff transformation and large S approximation is often to include remaining XY piece of the Hamiltonian and study effects of residual spin interactions (transverse quantum fluctuations) and obtain Goldstone modes. 

Presence of positive frequency quadratic/linear Goldstone modes for AFM, indicates emergence of long range order via spontaneous symmetry breaking. In the case of unfrustrated AFM's in dimensions 2 and above, transverse zero point spin fuctuations lead to (finite, non-divergent) zero point reduction in sublattice magnetization. Quantum fluctuations also modify Ising AFM (product) state into quantum entangled many body spin states.
\section*{Looking for\\Metastable Kitaev Spin Liquids in\\Isotropic Heisenberg Hamiltonians}

We observe that isotropic SU(2) symmetric Heisenberg Hamiltonian $H^H$, with nearest neighbor $\langle ij \rangle$ coupling in a honeycomb lattice can be symmetrically decomposed exactly, into three noncommuting pieces of Kitaev Hamiltonians,
\be
H^H = J\sum_{\langle ij \rangle} {\vec{\sigma}}_{i} \cdot {\vec{\sigma}}_{j}
~\equiv~ H^{K}_{xyz} + H^{K}_{yzx} + H^{K}_{zxy}
\ee
where the reference Kitaev piece
\be
H^K_{xyz} \equiv J\sum_{\langle ij \rangle_x} {\sigma}^{x}_{i}{\sigma}^{x}_{j} + J \sum_{\langle ij \rangle_y} {\sigma}^{y}_{i}{\sigma}^{y}_{j}+J \sum_{\langle ij \rangle_z} {\sigma}^{z}_{i}{\sigma}^{z}_{j}
\ee
The three noncommuting Kitaev pieces are related by cyclic permutations of spin components along 3 nearest neighbor bonds, denoted as ${\langle ij \rangle}_x, {\langle ij \rangle}_y$ and ${\langle ij \rangle}_z$ of the reference piece $H^K_{xyz}$ (equation 3).

In the supplementary section we summarise exact solution of manybody spectrum \cite{KitaevHC} of the reference Kitaev Hamiltonian. Kitaev writes Pauli spin operators $\sigma^{x}_i=ic^{x}_ic_i$, $\sigma^{y}_i=ic^{y}_ic_i$ and $\sigma^{z}_i=ic^{z}_ic_i$, using four Majorana Fermion operators $(c^x_i, c^y_i, c^z_i, c_i)\equiv ({\vec c}_i, c_i)$ (vector and a scalar), of Hilbert space dimension 4. This representation preserves commutation relation between components of the Pauli spin operators. The condition $\sigma^{x}_i \sigma^{y}_i \sigma^{z}_i = i $ for each site, imposes a local constraint $c^{x}_i c^{y}_i c^{z} c_i = 1$, and reduces dimension of the 4-dimensional Hilbert space of Majorana Fermion modes into 2-dimensional physical Hilbert space of Pauli spin operators.

In what follows we pick a reference Kitaev Hamiltonian piece $H^K_{xyz}(0)$, in the \textit{zero flux sector} - it is the Hamiltonian of free emergent Majorana Fermions hopping on the honeycomb lattice (equation 16, supplementary section), in the absence of flux (B$_p$ = 1) in every plaquette. Then we study, using certain approximations, how this reference metastable Kitaev liquid vacuum and its low energy excitation spectra are modified by interactions among Majorana Fermions, arising from interaction terms of remaining two Kitaev pieces.

As discussed in the next section, in the renormalized vacuum sector, Majorana Fermion hopping parameter gets modified and new vector Majorana Fermion collective modes emerge. All excitations have frequencies greater than or equal to zero. This qualifies the reference zero flux Kitaev spin liquid state as a metastable or locally stable phase, within our approximations.

\section*{Interactions Modified\\Metastable Kitaev spin liquids in\\isotropic Heisenberg Magnets}

In the Majorana Fermion representation for spin-\half operators Heisenberg Hamiltonian has the manifest rotational invariance, with respect to global SO(3) rotation of the vectors ${\vec c}_i$'s:
\be
H = J\sum_{\langle ij \rangle} {\vec{\sigma}}_{i} \cdot {\vec{\sigma}}_{j} = J \sum_{\langle ij \rangle} c_i c_j ( {\vec c}_i \cdot {\vec c}_j)
\ee
Scalar and vector Majorana Fermions get delocalized over the honeycomb lattice, via isotropic 2-body interactions. However, this 2-body interaction prevents exact solvability of SU(2) symmetric Heisenberg Hamiltonian, even in the enlarged Hilbert space. 

With a focuss on building metastable vacuua, we proceed to find the effect of the two remaining Kitaev pieces on the ground state Kitaev spin liquid sector of the reference Kitaev piece, using a Hartree type of approximate analysis. We introduce Heisenberg Hamiltonian H$^H$(0;xyz), for the ground state (zero flux) sector and take care of quantum fluctuations about the zero flux sector of reference Kitaev spin liquid (described by $ H^K_{xyz}(0)$), arising from interaction among Majorana Fermions, present in the remaining two Kitae pieces $H^{K}_{yzx} + H^{K}_{zxy}$. 

The effective Heisenberg Hamiltonian H$^H$(0;xyz), corresponding to reference zero flux sector acquires a specific anisotropic two body Majorana Fermion interaction form:
\be
H^H(0;xyz) = H^K_{xyz}(0) + H^{K}_{yzx} + H^{K}_{zxy}~~,
\ee
where
\bearr
H^K_{xyz}(0) &\equiv& 
    J \sum_{\langle ij \rangle} ic_ic_j {~~~\rm and} \\
H^{K}_{yzx} + H^{K}_{zxy} &\equiv&
     + J \sum_{\langle ij \rangle_x} c_{i}c_{j} (c^{y}_{i}c^{y}_{j} + c^{z}_{i}c^{z}_{j}) + \nonumber\\ 
     + J \sum_{\langle ij \rangle_y} c_{i}c_{j} (c^{z}_{i}c^{z}_{j} &+& c^{x}_{i}c^{x}_{j}) +
     J \sum_{\langle ij \rangle_z} c_{i}c_{j} (c^{x}_{i}c^{x}_{j} + c^{y}_{i}c^{y}_{j})~~
\eearr

In H$^H$(0;xyz), corresponding to zero flux sector, scalar Fermion has an independent dynamics (first term, from reference Kitaev piece). Vector Fermions have no indpendent dynamics. Scalar Majorana Fermion has bond dependent 2-body interactions with the vector Fermions (arising from remaining two Kitaev pieces, equation 7), which help delocalizes vector Fermions in an anisotropic fashion. This is to be contrasted with the full isotropic Heisenberg Hamiltonian (equation 4), where scalar and vector Fermions get delocalized in an isotropic fashion, via 2-body interactions. 

Now we perform a uniform Hartree factorization of interaction terms (equation 6) and get Hartree appoximated Heisenberg Hamiltonian ${H}^H_H(0;xyz)$ in the zero flux sector:
\bearr
{H}^H_H(0;xyz)
     =(J + \alpha_v) \sum_{\langle ij \rangle} ic_ic_j&+&\\ \nonumber
     +(J+\alpha_s) \{\sum_{\langle ij \rangle_x}i(c^{y}_{i}c^{y}_{j} + c^{z}_{i}c^{z}_{j})
     &+&\sum_{\langle ij \rangle_y}i(c^{z}_{i}c^{z}_{j} + c^{x}_{i}c^{x}_{j}) +\nonumber\\
     +\sum_{\langle ij \rangle_z}i(c^{x}_{i}c^{x}_{j} + c^{y}_{i}c^{y}_{j})\} + {\rm constant,}
\eearr
where the Hartree averages, $\alpha_v \equiv \frac{2}{3}\langle {\vec c}_i\cdot{\vec c}_j\rangle$ and $\alpha_s \equiv \langle {c}_i c_j\rangle$ are determined selfconsistently. We define renormalized hopping parameters, J$_v \equiv J(1+\alpha_v)$ for vector and J$_s \equiv J(1+\alpha_s)$ for scalar Fermions. 
After regrouping, above Hamiltonian takes a more transparent form:
\bearr  
{H}^H_H(G,xyz) = J_s \sum_{\langle ij \rangle} ic_ic_j &+& \nonumber\\
      +J_v\{\sum_{\langle ij \rangle_{yz}}ic^{x}_{i}c^{x}_{j}
     +\sum_{\langle ij \rangle_{zx}}ic^{y}_{i}c^{y}_{j}&+&
     \sum_{\langle ij \rangle_{xy}}ic^{z}_{i}c^{z}_{j}\}, ~~~~
\eearr
where the new bond symbol $\langle ij \rangle_{xy}$ stands for nearest neighbor hopping of z-component of vector Majorana Fermions, along a system of parallel xy zig-zag chains, made of alternating x and y type of bonds. Similarly x and y components of vector spins hop along yz and zx zig-zag chain systems respectively. While scalar Fermions continue to hop isotropically in the honeycomb lattice, vector Fermions acquire anisotropic 1D chain dynamics. That is, (x,y,z) components of vector Majorana Fermions become one dimensional and hop independently only along (yz, zx, xy) system of zig-zag chains respectively.

Using translational invariance and periodic boundary condition, we get Bloch states for scalar Fermion with 2D (graphene like) Kitaev spectrum and 1D spectrum for each components of vector Fermions, which hop along yz, zx and xy zig-zag chains. All energy eigen values are $\ge$ 0. This ensures that Kitaev spin liquid state is not completely destabilized by \textit{transverse quantum fluctuations} (present in the isotropic Heisenberg Hamiltonian) and survives as a metastable one.  Scalar Fermion spectrum is renormalized. While the scalar Majorana Fermions have Dirac nodes at K and K' points in, three components of vector Fermions have three independent sets of zero energy lines in the BZ. We have Majorana \textit{Fermi surfaces} in the 2D BZ, somewhat similar to Kitaev model in the Fisher lattice \cite{GBSantoshShankar}.

\section*{Spontaneous 2 spin Nematic Metastable order in Isotropic Heisenberg Magnets}

Metastable Kitaev spin liquid phase is anisotropic, as evidenced from 1D nature of the three components of vector Majorana Fermions, delocalized independently along three different sets of zig-zag chains. There is a nematic spin liquid order arising via symmetry breaking (from choice of reference Kitaev piece) in an SU(2) symmetric Heisenberg spin system. Vector Fermions, which were organized as static Z$_2$ gauge field in Kitaev model, organize themselves as positive frequency vector Fermion collective modes, in our isotropic Heisenberg Hamiltonian.

We obtain a manifold of Kitaev spin liquid metastable vacuua generated by SO(3) rotation. This is achieved by global rotation of spin axes of the reference Kitaev piece. The vector Fermionic modes we have obtained are collective modes, which represent generalized Goldstone modes. Interestingly, the Kitaev spin liquid metastable vacua exist for both Ferro or Antiferromagnetic isotropic Heisenberg systems (for J positive and negative).

Kitaev Hamiltonian is known to have highly anisotropic and ultra short range two spin correlation functions \cite{GBShankarMandal}, consistent with the anisotropic bond dependent Ising spin-spin interaction in Kitaev spin Hamiltonian. In particular,
\bearr
&&\langle \sigma_i^\alpha \sigma_j^\beta\rangle = g_{\langle ij \rangle_\alpha} \delta_{\alpha\beta}~~
~~{\rm i,j~nearest~neighbors}\nonumber\\&&{\rm ~~~~~~~~~=~0~~~~~~~~~~~~~~otherwise}
\eearr
In our approximation we find that bond dependent anisotropy gets modified only quantitatively, by quantum fluctuations contained in equation 7. An accurate and quantitative estimate of correction to spin-spin correlation function (unlike the case of pure Kitaev model \cite{GBShankarMandal}) involves projection of the Majorana Fermion wave function into physical spin Hilbert space, which we are not performing in the present article.

From the point of view of full rotational spin symmetry of isotropic Heisenberg Hamiltonian, above anisotropic spin-spin correlations represent emergent two spin nematic order present in the metastable Kitaev spin liquid we have constructed. It is a three sublattice (3 types of nearest neighbor bonds) nematic order (represented by three \textit{directors}, headless vectors), lying parallel to x,y and z directions in spin space.

It is also important to remember that eventhough we know approximate manybody spectrum, the manybody wave functions, including the ground state is not known in the physical Hilbert space, interms of physical spin variables.
 
\section*{Crafting by Symmetric Decomposition of Hamiltonians}
Often our focuss is on low energy manybody spectrum and low energy physics of model Hamiltonians. However, these Hamiltonians also organize a rich spectrum of excited manybody states and metastable phases in the Hilbert space. We can get information about metastable phases, when models are completely integrable or use exact diagonalization study of small systems, quantum Monte Carlo simulations, time dependent studies etc. One could also use renormalization group study, in multiparameter space of Hamiltonians, to explore metastable phases by a study of stable and unstable fixed points and flow in the parameter space.

We saw that symmetric splitting of unfrustrated isotropic Heisenberg Hamiltonians on the honeycomb lattice, into three noncommuting bond anisotropic (frustrated) Kitaev spin Hamiltonians (frustration arising from bond dependent anisotropic spin-spin interactions), helped us to find Kitaev spin liquid like metastable phases and associated spontaneous nematic order etc. This suggests an alternative approach \textit{symmetric decomposition of Hamiltonian, using symmetries of subgroups of full lattice translational and spin rotational symmetries}. 

Our hypothesis is that residual symmetries in the decomposed Hamiltonian protects corresponding metastable phases. Further, if the decomposed reference Hamiltonian piece is frustrated enough (like in Kitaev Hamiltonian) in spin space, it also ensures spin liquid metastable states.
 
There are multiple ways for Hamiltonian decomposition, where the pieces have symmetries of subgroups of lattice translation and spin rotation group. To illustrate this we introduce complementary Kitaev model $ H^{cK}_{xyz} \equiv H^K_{yzx}+ H^K_{zxy}$, such that isotropic Heisenberg Hamiltonian, is a sum of Kitaev nd complementary Kitaev Hamiltonians: $H^{H} \equiv H^K_{xyz} + H^{cK}_{xyz}$. We wish to find if metastable vacuua (complementary Kitaev spin liquid) corresponding to complementary Kitaev model exists. Since complementary Kitaev model is not exactly solvable, answering this question becomes difficult.

Let us consider a simpler situation of splitting unfrustrated isotropic spin-\half 2D Antiferromagnetic Heisenberg Hamiltonian on a triangular lattice, into Ising and XY pieces. We ask, is there is a manifold of metastable states where we have spontaneous Kosterlitz-Thouless XY order and corresponding Goldstone modes ?
 
Similarly, spin-\half XY model Hamiltonian in a square lattice H$^{XY}$ can be symmetrically decomposed into noncommuting sum of two Kugel-Khomskii \cite{KK} pieces $H^{KK}_{xy} {\rm and} H^{KK}_{yx}$: 
\be
H^{XY} = J\sum_{\langle ij \rangle} (\sigma^x_i\sigma^x_j + \sigma^y_i\sigma^y_j)
~\equiv~ H^{KK}_{xy} + H^{KK}_{yx}
\ee
Approximate analysis can be performed to find if square lattice XY FM or AFM contains metastable anisotropic spin liquid vacuua corresponding to Kuge-Khomskii compass Hamiltonians. 

There is a hope that designer metastable phases could be found out using our Symmetric Decomposition of Hamiltonians.

Using a similar approach, by adding and substracting SU(2) symmetric, isotropic spin liquid stabilizing frustrating terms in the frustration free isotropic  Hamiltonians, we could explore RVB type of SU(2) symmetric quantum spin liquids, which have a rich variety of quantum orders and chirality orders.

FM/AFM Hamiltonians of other lattices, where Kitaev model can be defined and exactly solved are being investigated by us \cite{MandalGB} looking for metastable Kitaev spin liquids: Fisher lattice (having Majorana FS \cite{GBSantoshShankar}), decorated honey comb lattice exhibiting spontaneous Chiral symmetry breaking \cite{KivelsonYao}, non-Archemedian lattice \cite{StoneWales}, classical spin Kitaev model \cite{GBSenShankar} 3D Kitaev model \cite{MandalSundar}. There are indications that because the lattices are more complex than honeycomb lattice, Kitaev spin liquids have deeper metastable minimum.

\section*{Preparation of False Vacuum and watching its evolution}

\textit{Fate of the false vacuum} \cite{ColemanFFV} is an important issue, from cosmology to our present problem of quantum tunneling and decay of metastable Kitaev spin liquid vacuum of an isotropic Heisenberg spin system. Since we have Jackeli-Khaliullin-Kitaev Heisernberg \cite{JackeliKhaliullin} materials with dominant isotropic Ferro/AFM interactions, it will be wonderful to perform real experiments. Also try to understand existing experimental results and anomalies \cite{Trebst,Shivram} in the light of intervension of metastable Kitaev spin liquid phases and quasiparticles.

As a number of quantum processors have emerged  and experimental simulations \cite{AshvinQP} and computer simulations \cite{SubirCompSim} are being performed on certain manybody lattice Hamiltonians, we suggest the following protocol to address our problem: i) prepare, best possible Kitaev spin liquid ground state for a finite number of qubits, using a reference (FM/AFM) Kitaev spin Hamiltonian and ii)  use the full isotropic spin-half (FM/AFM) Hesenberg Hamiltonian and study time evolution of the prepared Kitaev spin liquid with a nematic order. 

What do we expect ? Our hypothesis is presence of three stages in time evolution. Since there is a finite energy gap for Z$_2$ flux excitaions, we expect, first,  renormalization of the Kitaev spin liquid state (analogue of growth of zero point spin fluctuations in an Ising AFM vacuum in 3 or 2D). After a time interval, the system will nucleate Z$_2$ fluxes, create a small density of Majorana Fermions as well as topological defects in the Kitaev nematic order, via quantum tunneling. Finally there will be a crossover to nucleation of real (FM/AFM) ground state bubbles and their growth. 

In a model study we find \cite{GBTXtl} a remarkable time crystal like local transient behaviour, a periodic oscillation of the state vector. In short, there may be many surprises in store, in real experiments and theoretical studies.

\section*{Discussion}

In the present article we have explored possibility of metastable spin liquid states in istropic spin-\half Heisenberg systems on a honeycomb lattice. In addition, we have suggested general ways to explore presence of metastable spin liquid states. While some of the metastable states may turn out to be locally stable in all directions, others may not be locally stable in all directions.

Our study also raises some interesting questions. If we get Kitaev spin liquid and spin nematics in isotropic systems, it allows for topological singularities in the spin nematic phase, in addition to plaquette flux excitations. It is likely these topological defects trap non-Abelian anyons as zero modes.

Taking advantage of known exact specrum of Kitaev honeycomb lattice Hamiltonian, we have restricted our search to study quantum spin liquids that do not have manifest global SU(2) symmetry. We also find, in our approximate study of isotropic Heisenberg magnets, presence of SU(2) symmetric metastable spin liquids, a la' RVB states. Under certain conditions \textit{time crystal} like transient and local metastable phases emerge. 
 
It is indeed exciting that beyond the comfort zone of equilibrium quantum statistical mechanics, rich new worlds await even in energy eigen states and near eigen (manybody wave packet) states of familiar and well studied Hamiltonians. From the point of view of performing quantum computation and related tasks, present work opens new avenues to \textit{explore and use exotic quasiparticles and exotic metastable states, which are hiding in nonexotic real systems.}

\section*{Acknowledgement}
I thank Francesco Ferrari, R. Shankar and S. Mandal for discussions; and thank Tony Leggett for his encouraging remarks. Hospitality and stimulating visiting positions, at The Institute of Mathematical Sciences, Chennai, Indian Institute of Technology, Chennai and Perimeter Institute for Theoretical Physics (PI), Waterloo, Canada are acknowledged. PI is supported by the Government of Canada through Industry Canada and by the Province of Ontario through the Ministry of Research and Innovation.

\section*{Supplementary section}
\section*{Exact Solution of Kitaev honeycomb lattice model - a Summary}
Since we are looking for metastable Kitaev spin liquid phases in isotropic Heisenberg Hamiltonians, we review Kitaev's method of finding exact manybody spectrum of the Kitaev anisotropic spin Hamiltonian in the honeycomb lattce:
\be
H^K_{xyz} \equiv J\sum_{\langle ij \rangle_x} {\sigma}^{x}_{i}{\sigma}^{x}_{j} + J \sum_{\langle ij \rangle_y} {\sigma}^{y}_{i}{\sigma}^{y}_{j}+J \sum_{\langle ij \rangle_z} {\sigma}^{z}_{i}{\sigma}^{z}_{j}
\ee
Here 3 nearest neighbor bonds, where two spin Ising ineraction term, xx or yy or zz  alone are present are denoted respectively by ${\langle ij \rangle}_x, {\langle ij \rangle}_y$ and ${\langle ij \rangle}_z$. 
 
We assume periodic boundary condition for the hexagonal lattice, having N unit cells and 2N sites. Each site carries a spin-\half moment. Kitaev found that plaquette spin operators (sites 1 to 6 denote nearest neighbor sites of a given plaquette):
\be
B_p \equiv 
{\sigma}^{z}_{1}{\sigma}^{x}_{2}{\sigma}^{y}_{3}{\sigma}^{z}_{4}{\sigma}^{x}_{5}{\sigma}^{y}_{6}~~~~, 
\ee
defined for each elementary plaquette commute among themlselves and with $H^K_{xyz}$. Since B$^2_p =  1$, eigen values of B$_p = \pm 1$. Consequently, energy eigenstates of Kitaev Hamiltonian get classified interms of eigen values of the set $\{B_p\}$.

To get the full manybody spectrum Kitaev introduces a Majorana Fermion representation for Pauli spin operators in an enlarged Hilbert space. Each site has four Majorana Fermion operators, $(c^x_i, c^y_i, c^z_i, c_i)\equiv ({\vec c}_i, c_i)$ (vector and a scalar), of dimension 4. Dimension of Majorana Fermion Hilbert space is 4$^{2N}$. Dimension of physical spin Hilbert space is 2$^{2N}$.  

The representation, $\sigma^{x}_i=ic^{x}_ic_i$, $\sigma^{y}_i=ic^{y}_ic_i$ and $\sigma^{z}_i=ic^{z}_ic_i$,  preserves commutation relation between components of the Pauli spin operators. The condition $\sigma^{x}_i \sigma^{y}_i \sigma^{z}_i = i $ for each site, imposes a local constraint $c^{x}_i c^{y}_i c^{z} c_i = 1$, and reduces dimension of the 4-dimensional Hilbert space of Majorana Fermion modes into 2-dimensional physical Hilbert space of Pauli spin operators. 

A specially designed anisotropic structure of the Kitaev Hamiltonian H$^K_{xyz}$ makes it exactly solvable. In a remarkable fasion, certain pairs of vector Fermion components appearing in H$^K_{xyz}$ organize themselves into \textit{static and commuting Hartree vector potential}, which are emergent static Z$_2$ gauge fields. One obtains free (noninteracting) scalar Fermions, which hop and delocalize, in the presence of emergent static Z$_2$ gauge fields:
\bearr
&&H^K_{xyz} = 
J \sum_{\langle ij \rangle_x} c_{i}c_{j} c^{x}_{i}c^{x}_{j} + 
J \sum_{\langle ij \rangle_y} c_{i}c_{j} c^{y}_{i}c^{y}_{j} + 
J \sum_{\langle ij \rangle_z} c_{i}c_{j} c^{z}_{i}c^{z}_{j} \nonumber\\
&&\equiv J \sum_{\langle ij \rangle_x} ic_{i}c_{j} {\hat{u}^x}_{\langle ij \rangle_x}+
J \sum_{\langle ij \rangle_y} ic_{i}c_{j} {\hat{u}^y}_{\langle ij \rangle_y}+
J \sum_{\langle ij \rangle_z} ic_{i}c_{j} {\hat{u}^z}_{\langle ij \rangle_z}\nonumber\\
~~
\eearr
Operators ${\hat u}^{\alpha}_{\langle ij \rangle_\alpha} \equiv -ic^{\alpha}_{i}c^{\alpha}_{j}$ ($\alpha$ = x,y,z) appearing above commute among themselves and with H$^K_{xyz}$: consequently they can be treated as c-numbers in solving the manybody Hamiltonian. Further, $({\hat u}^{\alpha}_{\langle ij \rangle_\alpha})^2 = 1$ implies that ${\hat u}^{\alpha}_{\langle ij \rangle_\alpha}$ = $\pm 1$. 

The transformation ${\hat u}^{\alpha}_{\langle ij \rangle_\alpha} \rightarrow \tau_i{\hat u}^{\alpha}_{\langle ij \rangle_\alpha}\tau_j$ and $c_i \rightarrow \tau_i c_i$ leaves Kitaev Hamiltonan invariant. Here $\tau_i \pm 1$. We have an emergent local Z$_2$ gauge invariance and u's become emergent Z$_2$ gauge fields: Consequently, enlarged Hilbert space of dimension 4$^{2N}$ breaks into 2$^{2N}$ gauge copies, each copy having identical energy spectrum of 2$^{2N}$ manybody energy eigen states.

Formally, the exact many body spectrum becomes that of emergent non-interacting Majorana Fermions hopping in the lattice in the presence of static ${\hat u}^{\alpha}_{\langle ij \rangle_\alpha}$ fields. Number of distinct Z$_2$ gauge flux configurations is 2$^N$, where N is the number of elementary hexagons.
Wilson loops that are gauge invariant operators on hexagens, measure gauge fluxes (0 or $\pi$), get identified with the B$_p$ operators (equation 3). That is, B$_p \equiv {\hat u}^{y}_{\langle 12 \rangle_y}{\hat u}^{z}_{\langle 23 \rangle_z}{\hat u}^{x}_{\langle 34 \rangle_x}{\hat u}^{y}_{\langle 45 \rangle_y}{\hat u}^{z}_{\langle 56 \rangle_z}{\hat u}^{x}_{\langle 61 \rangle_x}$. Eigen values $\pm 1$ of B$_p$ corresponds to 0 and $\pi$ fluxes of emergent Z$_2$ gauge fields.

It follows from Lieb theorem that ground state of Kitaev Hamiltonian is the lowest energy state of zero Z$_2$ flux, $\{B_p = 1\}$ sector. Hamiltonian H$^K_{xyz}(0)$ of the zero flux sector is obtained by putting all ${\hat u}^{\alpha}_{\langle ij \rangle_\alpha} = 1$ in equation (4):
\be
H^K_{xyz}(0) = J \sum_{\langle ij \rangle} ic_{i}c_{j}
\ee
That is, in the zero flux sector, free scalar Majorana fermions hop isotropically in the honeycomb lattice. In this sector the vector Majorana Fermion degree of freedom essentially disappear from the problem.

\end{document}